\documentclass[onecolumn, showpacs,amsmath,amssymb,superscriptaddress]{revtex4}
\usepackage{amsmath,amssymb,color,latexsym,natbib,graphicx}
\usepackage{dcolumn, enumerate}
\usepackage{bm}
\def\ADD#1{{\textcolor{red}{#1}}}    
 
\newcommand{\p} {\partial}

\def\uu{{\bf u}}

\def\d_M{{\bf d_M}}

\def\rr{{\bf r}}

\def\bb{{\bf b}}

\def\bOmega{{\boldsymbol \Omega}}
\def\bomega{{\boldsymbol \omega}}
\def\be{\begin{equation}}
\def\ee{\end{equation}}
\def\ba{\begin{eqnarray}}
\def\ea{\end{eqnarray}}
\def \pmbmath{\mathpalette\pmbmathaux}
\def \pmbmathaux#1#2{
	\pmbtext{$#1#2$}}
\def \pmbtext#1{\leavevmode
	\setbox0\hbox{#1}
	\kern0,4pt \copy0 \kern-\wd0
	\kern-0,2pt \raise0,3pt \box0 }

\usepackage{microtype}
\usepackage{calrsfs}
\DeclareMathAlphabet{\pazocal}{OMS}{zplm}{m}{n}
\usepackage[cal=rsfso,calscaled=.96]{mathalfa}
\DeclareMathAlphabet{\mathpzc}{OT1}{pzc}{m}{it}
\DeclareMathAlphabet{\mathcalligra}{T1}{calligra}{m}{n}

\newcommand{\mnras}{Mon. Not. R. Astron. Soc.}

\newcommand{\jfm}{J. Fluid Mech.}

\def\ADD#1{{\textcolor{black}{#1}}}    

\def\uu{{\bm u}}
\def\BB{{\bm B}}
\def\bb{{\bm b}}
\def\gg{{\bm g}}

\def\ag{{\bm {A_G}}}
\def\jm{{\bm {j}}}
\def\jb{{\bm {j_b}}}
\def\rr{{\bm r}}

\def\Nabla{{\pmbmath {\nabla}}}
\def\be{\begin{equation}}
\def\ee{\end{equation}}
\def\ba{\begin{eqnarray}}
\def\ea{\end{eqnarray}}
\begin{document}
	
	\preprint{1}
	
	\title{{Energy transfer in compressible magnetohydrodynamic turbulence for isothermal self-gravitating fluids}}
	\author{Supratik Banerjee}
	\email{supratik.banerjee@uni-koeln.de}
	\affiliation{Universit\"at zu K\"oln, Institut f\"ur Geophysik und Meteorologie, Pohligstrasse 3, 50969 K\"oln, Germany}
	\author{Alexei G. Kritsuk}
	\email{akritsuk@ucsd.edu}
	\affiliation{University of California, San Diego, 9500 Gilman Drive, La Jolla, California 92093-0424, USA}

	\date{\today}
	
	\begin{abstract}
		Three-dimensional, compressible, magnetohydrodynamic turbulence of an isothermal, self-gravitating fluid is analyzed using two-point statistics in the asymptotic limit of large Reynolds numbers (both kinetic and magnetic). Following an alternative formulation proposed by S. Banerjee and S. Galtier (Phys. Rev. E {\bf 93}, 033120, 2016) and S. Banerjee and S. Galtier (J. Phys. A, Math. and Theor. {\bf 50}, 015501, 2017), an exact relation has been derived for the total energy transfer. This approach results in a simpler relation expressed entirely in terms of mixed second-order structure functions. The kinetic, thermodynamic, magnetic and gravitational contributions to the energy transfer rate can be easily separated in the present form. By construction, the new formalism includes such additional effects as global rotation, the Hall term in the induction equation, etc. The analysis shows that solid-body rotation cannot alter the energy flux rate of compressible turbulence. However, the contribution of a uniform background magnetic field to the flux is shown to be non-trivial unlike in the incompressible case. Finally, the compressible, turbulent energy flux rate does not vanish completely due to simple alignments, which leads to a zero turbulent energy flux rate in the incompressible case.
		
	\end{abstract}
	\maketitle

	\section{Introduction}
Turbulence is a non-linear phenomenon omnipresent in nature. For a fully developed turbulence in the limit of infinitely large Reynolds numbers, the fluid flow contains fluctuations, populating a wide range of length and time scales. In the inertial range sufficiently decoupled from the large-scale forcing and small-scale dissipation, energy (and other inviscid invariants of motion) takes part in a cascade process, transporting it across scales.

During this process, the average flux rate $\varepsilon$ of an inviscid invariant is independent of the length scale, thereby characterizing a turbulent state. In the analytical framework of homogeneous turbulence, there exist a number of exact relations which express $\varepsilon$ in terms of the statistical average of two-point correlation functions or two-point differences of relevant variables of the flow field (e.g., fluid velocity, density, magnetic field, etc.). In three dimensions, this formalism provides an accurate quantitative estimate of the kinetic energy dissipation rate, and hence of the heating rate of a system by the process of turbulent cascade. For incompressible hydrodynamic (HD) and magnetohydrodynamic (MHD) turbulence, such exact relations \citep{Kolmogorov41, Monin75, Politano98} can express $\varepsilon$ purely in terms of two-point differences.
 However, the basic assumption of homogeneity and the existence of a so-called inertial zone cannot be guaranteed, in general, for compressible turbulence. Nevertheless, as the first step, the differential form derived by Monin \cite{monin59} was extended to compressible isothermal HD, MHD, and polytropic HD, considering the total energy as inviscid invariant \citep{Galtier11, Banerjee13, Andres17, Banerjee14}. The relation for isothermal HD turbulence \citep{Galtier11} was then numerically verified for a driven supersonic regime (at r.m.s. Mach number $\sim$ 6) \citep{Kritsuk13}. All these compressible exact relations were written in a flux-source form 
\begin{equation}
- 4 \varepsilon = \nabla \cdot {\pazocal F} + {\pazocal S} , 
\end{equation}
where the source ${\pazocal S}$ vanishes identically in the incompressible case, retaining only the divergence of the flux ${\pazocal F}$ on the right-hand side (rhs). Unlike the divergence part, the source included both single-point and two-point contributions, making the interpretation difficult. 
Moreover, unlike the incompressible exact relations, straightforward formulations of compressible counterparts were not unique, indicating that some essential physical constraints are missing. Recently, we derived an exact relation for self-gravitating, isothermal turbulence \citep{Banerjee17}, and showed that proper accounting for the acoustic energy equipartition eliminates single-point contributions from the source ${\pazocal S}$.  Furthermore, the new constraint implied that the correlation between the velocity and pressure dilatation does not actually play any role in the total energy cascade process in isothermal HD turbulence which was claimed previously \citep{Aluie11}. \ADD{ However, the previous flux-source formulation contained certain terms which could be cast neither as a pure divergence term (or a so-called flux term) nor as a source term. In addition, there were also certain terms which could be cast as both flux and source terms \citep{Andres17}. Finally, the compressible flux-source form is cumbersome and in an anisotropic case its practical applications are subject to non-trivial numerical integration.}

In this paper, we have generalised the alternative formulation of exact relation derived recently for incompressible HD and MHD turbulence \citep{Banerjee16, Banerjee17JoP} to the case of three-dimensional, self-gravitating, isothermal MHD turbulence, assuming statistical homogeneity. In contrast to the previous flux-source form, the current formulation casts the invariant flux rate $\varepsilon$ in terms of mixed second order structure functions, associating the fluctuations of the density ($\rho$), the fluid velocity ($\uu$), the vorticity ($\bomega$), the magnetic field ($\BB$), the current ($\nabla \times \BB$) and other variables composed by them.  Interestingly, the new form is easily amenable to the turbulence of a rotational fluid and also can be readily extended to the Hall MHD turbulence.
In addition, thanks to the new formulation, we have algebraic forms (free of a global divergence) for general anisotropic case, which will be of great interest for the studies of solar wind turbulence \ADD{(the turbulence is anisotropic due to a non-negligible background magnetic field). In the framework of solar wind turbulence, this type of exact relation is crucial to determine the turbulent heating rate \citep{Banerjee16apj, Hadid17} and also to understand the wind acceleration mechanisms \citep{Ballegooijen16}.} However, the main driver behind this work, quite naturally, is astrophysics of star formation \citep{Mckee07}. There, the importance of molecular cloud turbulence, rotation, magnetic, and gravitational effects was recognized a long time ago, while the lack of rigorous theoretical results hindered the progress, opening the field to competing speculative scenarios. One of the key unsolved problems is how exactly the nonlinear coupling of the turbulence with magnetism and gravity result in self-organized fluid dynamics within star-forming molecular clouds. \ADD{The rigorous approach taken in this contribution will help to shed light on that problem. In addition, this exact relation will help to understand the detailed energy equipartition of distinct modes in compressible MHD turbulence, and the role of shocks, which are crucial in compressible fluid flows.} 

The paper is structured as follows. In Sec. II we discuss the basic equation, the energy conservation and the explicit forms of two-point correlators of the total energy. 
The following section consists of the derivation of the exact relation along with its pure hydrodynamic version. In the same section, we also propose a shorter way to derive the final exact relation. Sec. IV is dedicated to the discussion of different features of the current exact relation and its advantages. Finally, Sec. V summarizes the whole paper.

\section{Basic Equations}
In this paper, we are interested in a self-gravitating isothermal magnetohydrodynamic fluid. The basic equations are given as 
\begin{align}
\p_t \rho + \pmbmath{\nabla} \cdot \jm  &= 0,  \label{continuity}\;\;\;\;\;\;\;\; \\
\partial_t \jm + \pmbmath{\nabla} \cdot (\jm \otimes \uu) &= - \pmbmath{\nabla} p +   \jb \times \bb  + \rho {\gg}  + {\bm d_H} + {\bm f}, \label{momentum} \\
\p_t \bb &= \nabla \times \left( \uu \times \bb \right) + {\bm d_M}, \label{induction} \\
\nabla \cdot {\gg} &= - 4 \pi G \left(\rho - \rho_0 \right),
\label{hd1}
\end{align}
where $ \bb = \BB/\sqrt{\mu_0}$ with $\BB$ being the magnetic field, $\jm= \rho \uu$, $\jb = \nabla \times \bb$ and $p = c_s^2 \rho$ with $c_s$ being the (constant) isothermal sound speed. ${\bm d_H}$, ${\bm d_M}$ and ${\bm f}$ represent the kinetic dissipation, magnetic dissipation and a stationary forcing. In the Poisson equation (\ref{hd1}), we use the density fluctuation with respect to the spatial average $\rho_0 $ (also the statistical average) rather than the local density $\rho$, which is compatible with periodic boundary conditions often used in simulations of turbulence in astrophysical systems. Now taking partial time derivative of the momentum conservation equation and also using the continuity equation, we get 
\begin{equation}
\Nabla \cdot   {\partial_t \gg}  =  4 \pi G \nabla \cdot (\rho \uu) =>  \partial_t \gg - 4 \pi G \rho\uu = \nabla \times {\ag},  \label{evolution}
\end{equation}
where $ \ag $ is the vector potential related to gravitation. 

The total energy density at any point of the flow field is equal to the sum of the densities of kinetic energy, magnetic energy, gravitational potential energy and thermodynamic energy at that point and can be written as 
\begin{equation}
{\pazocal E} =  \left(  \rho \bm u^2 + \bb^2 - \alpha \bm g^2 \right)/2 + \rho e  , \label{consE}
\end{equation}
where $\alpha  = 1/(4 \pi G)$ and $e = c_s^2\ln(\rho/\rho_0)$ is the thermodynamic potential energy.
Note that, the fluid velocity $\uu$ at every point describes the fluctuation velocity about its statistical mean value (which can be eliminated by a suitable Galilean transformation) whereas the density, the internal energy, the magnetic field and the gravity field corresponds to the total values at every point.

To show the conservation of energy when the viscous and forcing terms are neglected, 
we evaluate the time derivative of each term of total energy. We have (assuming periodic boundary conditions or zero velocity on the boundary surface)
\begin{align}
  \int_V {\partial_t} \left(\frac{\rho \uu^2}{2}\right) d\bm x   &= \int_V \left[ \jm \cdot \gg - \uu \cdot \nabla p + \uu \cdot \left( \jb  \times \bb \right)  \right]  d\bm x,  \\
 \int_V {\partial_t} \left(\frac{\bb^2}{2}\right) d\bm x   &=   \int_V \jb \cdot \left( \uu \times \bb \right)   d\bm x  =  -  \int_V \uu \cdot \left( \jb \times \bb \right)   d\bm x,  \\
- \frac{\alpha }{2}\int_V {\partial_t} \gg^2 d\bm x  &= \int_V \left[ \nabla\! \cdot\! \left( \gg\! \times\! {\bm a_G} \right) - \jm \! \cdot\! \gg\right]  d\bm x, \\
\int_V {\partial_t (\rho e)} d\bm x &= \int_V   \uu \cdot \nabla  p\, d\bm x, 
\end{align}
where ${\bm  a}_G = \alpha{\bm A}_G/2$.
Adding up the three above expressions and assuming additionally that the boundary surface is either periodic or gravitationally equipotential ($\gg=0$), 
we can prove the total energy conservation for a non-viscous system without external forcing.  From the above equations, it is evident that the kinetic energy exchanges with each of the three types of potential energy (magnetic, gravitational and thermodynamic) whereas the three components of potential energy evolve independently of each other. In fact, for such a fluid, one can show that there is equipartition between the average kinetic energy and the average total potential energy \citep{Zweibel95}. This type of equipartition also holds for the linear wave modes of compressible MHD. Following the same argument of \cite{Banerjee17}, we can define the two-point symmetric correlator of total energy in the current case as 
\begin{equation}
{\pazocal R}(\rr) = \left\langle \frac{R_\pazocal{E} + R'_\pazocal{E}}{2} \right\rangle \label{14}
\end{equation}
with
\begin{equation}
R_{\pazocal E} \equiv (\jm  \cdot {\uu}' + \bb \cdot \bb' - \alpha {\gg} \cdot {\gg}' + {\rho} {e}' + \rho e )/2   \quad  \text{and}  \quad  R'_{\pazocal E} \equiv (\jm' \cdot {\uu} + \bb' \cdot \bb - \alpha {\gg}' \cdot {\gg} +{\rho}' {e} + \rho'e')/2. \label{ecf}
\end{equation}

\section{Derivation of the exact relation} 

Unlike the exact relations for compressible isothermal MHD turbulence which have been derived \citep{Banerjee13, Andres17}, in this paper we propose an alternative form which is an extension of \citep{Banerjee17JoP} in case of compressible fluids. Under this formalism, the final extact relation is simpler and hence for a physical system, the calculation of the energy flux rate becomes much easier. To derive the alternative form, we use the Lamb formulation of Navier-Stokes equations. Hence the equation (\ref{momentum}) can be re-written as 
\begin{align}
\partial_t \jm &= - \uu \left( \nabla \cdot \jm \right) - \rho \nabla \left( \frac{u^2}{2} \right) +  \left(\jm  \times \pmbmath{\omega} \right) - \pmbmath{\nabla} p +  \left( \jb \times \bb \right) + \rho {\gg} + {\bm d_H} + {\bm f}  ,  \, \, \text {and} \\
\partial_t \uu &= -  \nabla \left( e + \frac{u^2}{2} \right) + \left( \uu \times \pmbmath{\omega} \right) +  \frac{ \left( \jb \times \bb \right)}{\rho} + \gg + \frac{1}{\rho} ({\bm d_H} + {\bm f}) .
\end{align}

The next step is to derive the evolution equations for ${\pazocal R}(\bm r)$. Using equations (\ref{continuity}), (\ref{momentum}), (\ref{induction}), (\ref{hd1}), we can write 
\begin{align}
\p_t \left\langle \jm \cdot \uu' \right\rangle &=   \langle \jm \cdot \p_t \uu' + \uu' \cdot \p_t \jm \rangle \nonumber\\
&=  \left\langle  \jm \cdot \left[ - \Nabla' \left( e'  + \frac{u'^2}{2}\right) + \left( \uu' \times \pmbmath{\omega}' \right) + \frac{ \left(\jb' \times \bb' \right)}{\rho'} + \gg' +  \frac{1}{\rho'} ({\bm d'_H} + {\bm f'}) \right]  \right\rangle \nonumber \\
&+ \left\langle  \uu' \cdot \left[  - \uu \left( \nabla \cdot \jm \right) - \rho \nabla \left( \frac{u^2}{2} \right) +  \left(\jm  \times \pmbmath{\omega} \right) - \pmbmath{\nabla} p +  \left( \jb \times \bb \right) + \rho {\gg} +  ({\bm d_H} + {\bm f}) \right]  \right\rangle,  \\
\p_t \left\langle {\bb} \cdot {\bb}' \right\rangle &=  \langle  \bb \cdot \p_t \bb' + \bb' \cdot \p_t  \bb \rangle \nonumber\\
&= \langle  \bb \cdot  [ \nabla' \times \left( \uu' \times \bb' \right) + {\bm d'_M} ] +  \bb' \cdot   \nabla \times [ \left( \uu \times \bb \right) + {\bm d_M} ]  \rangle \nonumber\\
&=    \langle \jb \cdot \left( \uu' \times \bb' \right) +  + \bb \cdot {\bf d'}_M  +  \jb' \cdot \left( \uu \times \bb \right)  + \bb' \cdot {\bf d}_M \rangle ,     \\
\p_t \left\langle {\gg} \cdot {\gg}' \right\rangle &= \langle  \gg \cdot \p_t \gg' + \gg' \cdot \p_t  \gg \rangle \nonumber\\
&=  \langle \Nabla \cdot (\ag \times \gg') + \Nabla' (\ag' \times \gg) \rangle + 4 \pi G \langle \jm \cdot  \gg'  + \jm'  \cdot  \gg  \rangle \nonumber\\
&=  \alpha^{-1} \left\langle \jm \cdot  \gg'  + \jm'  \cdot  \gg  \right\rangle , \\
\p_t \langle {\rho} {e}' \rangle &= \langle \rho \ \p_t e' + e' \ \p_t \rho \rangle =  \left\langle \jm\cdot\Nabla'e' - p \theta' - \rho\uu'\cdot\Nabla'e'\right\rangle ,  \\
\p_t \langle \rho e \rangle &= \p_t \langle \rho' e' \rangle = \langle \rho \ \p_t e + e \ \p_t \rho \rangle = - \left\langle p \theta \right\rangle , 
\end{align}
where $ \theta \equiv \nabla \cdot \uu$ and  we have used the statistical homogeneity to obtain $ \langle  \Nabla \cdot (\ag \times \gg') \rangle = - \langle  \Nabla' \cdot (\ag \times \gg')\rangle = \langle \ag \cdot (\Nabla' \times \gg') \rangle =0 $. Finally, we calculate the evolution of ${\pazocal R}$, which gives
\begin{align}
\p_t {\pazocal R} &= \frac{1}{4} \p_t \left \langle \jm \cdot \uu' + \jm' \cdot \uu + 2 \bb \cdot \bb' - 2 \alpha \gg \cdot \gg' + \rho e ' + \rho e + \rho' e + \rho' e' \right\rangle \nonumber\\
&= \frac{1}{4} \left\langle - (\uu' \cdot \uu) (\nabla \cdot \jm) - \rho \nabla \left(\frac{u^2}{2} \right) \cdot \uu' + \uu' \cdot (\jm \times {\bomega}) - \uu' \cdot \nabla p + \uu' \cdot ( \jb \times \bb) + \uu' \cdot \rho \gg + \uu' \cdot ({\bm d_H} + {\bm f}) \right.  \nonumber \\
& \left. - \jm  \cdot \nabla' \left( e' + \frac{u'^2}{2} \right) + \jm \cdot ( \uu' \times \bomega' ) + \jm \cdot \frac{(\jb' \times \bb')}{\rho'} + \jm \cdot \gg' + \frac{\jm}{\rho'} \cdot ({\bm d'_H} + {\bm f'})  \right. \nonumber \\
& \left. - (\uu \cdot \uu') (\nabla' \cdot \jm') - \rho' \nabla' \left(\frac{u'^2}{2} \right) \cdot \uu + \uu \cdot (\jm' \times {\bomega'}) - \uu \cdot \nabla' p' + \uu \cdot ( \jb' \times \bb') + \uu  \cdot \rho' \gg'  + \uu \cdot ({\bm d'_H} + {\bm f'})\right. \nonumber \\
& \left.  - \jm'  \cdot \nabla \left( e + \frac{u^2}{2} \right) + \jm' \cdot ( \uu \times \bomega ) + \jm' \cdot \frac{(\jb \times \bb)}{\rho} + \jm' \cdot \gg + \frac{\jm'}{\rho} \cdot ({\bm d_H} + {\bm f}) \right. \nonumber \\
& \left. + 2 \jb \cdot (\uu' \times \bb') + 2 \bb  \cdot {\bm d'_M}  + 2 \jb' \cdot (\uu \times \bb) + 2 \bb'  \cdot {\bm d_M} - 2 \jm \cdot \gg' - 2 \jm' \cdot \gg \right.  \nonumber \\
& \left. + \jm \cdot \nabla' e' - p \theta' - \rho \uu' \cdot \nabla' e' - p \theta + 
\jm' \cdot \nabla e - p' \theta - \rho' \uu \cdot \nabla e - p' \theta' \vphantom{\frac{1}{2}} \right \rangle.
 \end{align}
Equivalently, one can write 
\begin{equation}
\p_t {\pazocal R} = {\pazocal K} + {\pazocal M} + {\pazocal W} + {\pazocal U} + {\pazocal D} + {\pazocal F}
\end{equation}
where ${\pazocal K}$,  ${\pazocal M}$, ${\pazocal W}$, ${\pazocal U}$ account for, respectively, the kinetic, magnetic, gravitational, and thermodynamic contributions for the energy correlator. ${\pazocal D}$ and ${\pazocal F}$ denote, respectively, the total dissipative and forcing terms. All the terms can be explicitly written as
\begin{align}
{\pazocal K} &= -\frac{1}{4} \left\langle (\uu' \cdot \uu) (\nabla \cdot \jm) + \rho \nabla \left(\frac{u^2}{2} \right) \cdot \uu' - \uu' \cdot (\jm \times {\bomega})   + \jm  \cdot \nabla' \left( \frac{u'^2}{2} \right) - \jm \cdot ( \uu' \times \bomega' ) \right. \nonumber \\
& \left.  (\uu \cdot \uu') (\nabla' \cdot \jm') + \rho' \nabla' \left(\frac{u'^2}{2} \right) \cdot \uu - \uu \cdot (\jm' \times {\bomega'}) + \jm'  \cdot \nabla \left( \frac{u^2}{2} \right) - \jm' \cdot ( \uu \times \bomega ) \right \rangle , \\
{\pazocal M} &= \frac{1}{4} \left\langle  \uu \cdot ( \jb' \times \bb')+ \uu' \cdot ( \jb \times \bb)+ \jm \cdot \frac{\jb' \times \bb'}{\rho'}  +  \jm' \cdot \frac{\jb \times \bb}{\rho} + 2 \jb \cdot (\uu' \times \bb') + 2 \jb' \cdot (\uu \times \bb)    \right \rangle,     \\
{\pazocal W} &= -\frac{1}{4} \left\langle \jm \cdot \gg' + \jm' \cdot \gg -\rho \gg\cdot\uu'  - \rho' \gg' \cdot\uu   \right \rangle , \\
{\pazocal U} &=  -\frac{1}{4} \left\langle \rho \uu' \cdot \nabla' e' + \rho' \uu \cdot \nabla e  + p \theta + p' \theta' \right \rangle , \\
{\pazocal D} &= \frac{1}{4} \left\langle  \uu' \cdot {\bm d_H} +  {\jm} \cdot \frac{\bm d'_H}{\rho'}  +  \uu \cdot {\bm d'_H} +  {\jm'} \cdot \frac{\bm d_H}{\rho} + 2 \bb  \cdot {\bm d'_M} + 2 \bb'  \cdot {\bm d_M}\right \rangle , \\
{\pazocal F}  &= \frac{1}{4} \left\langle  \uu' \cdot {\bm f} +  {\jm} \cdot \frac{ {\bm f'} }{\rho'}  +  \uu \cdot {\bm f'}  +  {\jm'} \cdot \frac{\bm f}{\rho} \right \rangle.
\end{align}

Now, by straightforward algebra, one can show  following (Banerjee and Galtier, JoP, 2017) that 
\begin{align}
{\pazocal K} &=   \frac{1}{4} \left\langle  \delta \jm   \cdot   \delta \left[ \left( \uu \cdot \nabla \right) \uu \right] + \delta \uu \cdot  \delta \left[ \nabla \cdot \left( \jm \otimes \uu \right) \right] \right\rangle , \\
{\pazocal M} &=  \frac{1}{4} \left\langle  \delta \jm \cdot \delta \left( \frac{\bb \times \jb}{\rho} \right) + \delta \uu \cdot \delta \left( \bb \times \jb \right) + 2 \delta \jb \cdot\delta \left( \bb \times \uu \right) \right\rangle , \\
{\pazocal W} &= \frac{1}{4} \left\langle \delta \jm \cdot \delta \gg - \delta \uu \cdot \delta \left( \rho \gg \right)  \right \rangle ,  \\
{\pazocal U} &=  \frac{1}{4} \left\langle \delta \rho \,  \delta \left[ \left( \uu \cdot \nabla \right) e \right] \right \rangle ,
\end{align}
where we use the following algebraic manipulations:
\begin{enumerate}[(i)]
	\item $
	\left\langle \uu' \cdot (\jm \times {\bomega}) +  \jm \cdot ( \uu' \times \bomega' ) + \uu \cdot (\jm' \times {\bomega}') +  \jm' \cdot ( \uu \times \bomega )  \right\rangle  = - \left\langle \delta \jm \cdot \delta \left( \uu \times \bomega \right) + \delta \uu \cdot \delta \left( \jm \times \bomega \right) \right\rangle$, 
	
	where we use the fact that $ \jm \cdot \left( \uu \times \bomega \right) = -  \uu \cdot \left( \jm \times \bomega \right) = 0 $;
	
	\item $\left\langle - (\uu' \cdot \uu) (\nabla \cdot \jm) - \rho \nabla \left(\frac{u^2}{2} \right) \cdot \uu'  - \jm  \cdot \nabla' \left( \frac{u'^2}{2} \right) - (\uu \cdot \uu') (\nabla' \cdot \jm') - \rho' \nabla' \left(\frac{u'^2}{2} \right) \cdot \uu  - \jm'  \cdot \nabla \left( \frac{u^2}{2} \right)  \right \rangle $ \\
	$= \left\langle  \delta \uu \cdot \delta \left[ \left( \nabla \cdot \jm \right) \uu + \rho \nabla \left(\frac{u^2}{2} \right) \right] + \delta \jm \cdot \delta \left[\nabla \left( \frac{u^2}{2} \right) \right] \right\rangle $,
	
	where we used the fact that $\left\langle \left( \uu \cdot \uu \right) \left( \nabla \cdot \jm \right) + 2 \jm \cdot \nabla \left(\frac{u^2}{2} \right) \right\rangle = \left\langle \nabla \cdot \left(u^2 \jm \right)  \right\rangle = 0$ (by Gauss' divergence theorem) for homogeneous turbulence;  
	
	\item $\left\langle \jm \cdot \left( \frac{\bb \times \jb}{\rho} \right) + \uu \cdot \left( \bb \times \jb \right) + 2 \jb \cdot \left( \bb \times \uu \right) \right\rangle = 0$;  and finally
	
	\item $\left\langle \jm \cdot \nabla e + p \left( \nabla \cdot \uu \right) \right\rangle = \left\langle \uu \cdot \nabla p +  p \left( \nabla \cdot \uu \right)  \right\rangle =   \left\langle \nabla \cdot \left( p \uu \right) \right\rangle = 0 $.
	
\end{enumerate}

Now for a stationary state where the average total energy of the system does not change with time, the time derivative of the total energy correlators can also be made to vanish  due to the local isotropy hypothesis of Kolmogorov \citep{Lindborg96}. In addition, considering the length scales very far from the dissipative scale, we can neglect all the dissipative contributions and hence the final exact relation takes the form 

\begin{align}
- 4 \varepsilon &= \left\langle  \delta \jm   \cdot   \delta \left[ \left( \uu \cdot \nabla \right) \uu +   \frac{\bb \times \jb}{\rho}  + \gg \right] + \delta \uu \cdot  \delta \left[ \nabla \cdot \left( \jm \otimes \uu \right) 
+  \bb \times \jb  - \rho \gg  \right]  \right\rangle  \nonumber \\
&+  \left\langle  2 \delta \jb\cdot \delta \left( \bb \times \uu \right)   + \delta \rho \,  \delta \left[ \left( \uu \cdot \nabla \right) e \right] \right \rangle .\label{mainexact}
\end{align} 
where $\pazocal F = \varepsilon$ gives the net energy flux rate of the flow.

Equation (\ref{mainexact}) is the principal result of this paper. This describes an exact relation of three-dimensional, self-gravitating, compressible turbulence of an isothermal MHD fluid. Equation (\ref{mainexact}) expresses the flux rate of total energy purely in terms of two-point spatial differences. In the pure HD case of self-gravitating isothermal fluid, $\bb$ and $\jb$ vanish and hence the resulting exact law takes the form 
\begin{equation}
- 4 \varepsilon = \left\langle  \delta \jm   \cdot   \delta \left[ \left( \uu \cdot \nabla \right) \uu + \gg \right] + \delta \uu \cdot  \delta \left[ \nabla \cdot \left( \jm \otimes \uu \right) 
 - \rho \gg  \right]  + \delta \rho \,  \delta \left[ \left( \uu \cdot \nabla \right) e \right] \right \rangle, \label{mainhydro}
\end{equation} 
which has also a simpler and more compact form than that obtained in Ref.~\citep{Banerjee17}. \\

\paragraph*{\textbf{Alternative way of derivation}}
The equation (\ref{mainexact}) can be derived in a slightly different way which is shorter than the method described above. By definition, one can write 
\begin{align}
\p_t {\pazocal R} = &\frac{1}{4} \p_t \left\langle \jm  \cdot {\uu}' + \jm' \cdot {\uu} + 2 \bb \cdot \bb' - 2 \alpha {\gg} \cdot {\gg}' + {\rho} {e}' + \rho e +{\rho}' {e} + \rho'e'  \right\rangle \nonumber \\
=  &\frac{1}{4} \left\langle \p_t \jm  \cdot {\uu}' + \jm  \cdot \p_t {\uu}'  + \p_t \jm' \cdot {\uu} + \jm' \cdot \p_t {\uu} + 2 \p_t \bb \cdot \bb' + 2 \bb \cdot \p_t \bb' - 2 \alpha  \p_t {\gg} \cdot {\gg}' - 2 \alpha {\gg} \cdot \p_t {\gg}'  \right. \nonumber \\ 
& \, \, \left. +  {e}' \p_t{\rho} + {\rho} \p_t{e}' + \p_t (\rho e) + {e} \p_t{\rho}'  +  {\rho}' \p_t{e} + \p_t(\rho'e')  \right\rangle \, . \label{alter1}
\end{align}
In addition, the conservation of the total energy gives
\begin{equation}
\frac{1}{4}\p_t \left\langle \jm \cdot \uu + \jm' \cdot \uu' +  \bb \cdot \bb + \bb' \cdot \bb' -  \alpha {\gg} \cdot {\gg} - \alpha {\gg'} \cdot {\gg'} + 2 \rho e + 2 \rho'e'  \right\rangle = 0 . \label{alter2}
\end{equation}
Now subtracting (\ref{alter2}) from (\ref{alter1}), one can write 
\begin{equation}
\p_t {\pazocal R} = - \frac{1}{4} \left\langle \delta \jm \cdot \delta \left( \p_t \uu \right) + \delta \uu \cdot \delta \left( \p_t \jm \right) + 2  \delta \bb \cdot \left( \p_t \bb \right) - 2 \alpha \delta \gg \cdot \left( \p_t \gg \right) + \delta \rho \delta \left( \p_t e \right) + \delta e \delta \left( \p_t \rho \right)  \right\rangle.
\end{equation}
Now a mere substitution of the partial time derivatives and a little re-arrangement yields (\ref{mainexact}). However, using this approach, one does not have to explicitly calculate the contributions ${\pazocal K}$,  ${\pazocal M}$, ${\pazocal W}$, and ${\pazocal U}$, which contain the important information about the energy cascading process in spectral space (see \citep{Banerjee17}).

\section{Properties of the exact relation}

\paragraph{\textbf{Incompressible limit}}
For an incompressible fluid, at all points $\rho =\rho_0= \text{const.}$ and hence $e = 0$ and $\theta = 0$. If we further assume $\rho_0=1$,  (\ref{mainexact}) reduces to
\begin{equation}
- 4 \varepsilon = 2 \left\langle  \delta \uu   \cdot   \delta \left[ \left( \uu \cdot \nabla \right) \uu +   \left(\bb \times \jb \right)  \right] +  \delta \jb\cdot \delta \left( \bb \times \uu \right)  \right \rangle \, . \label{mainincomp}
\end{equation} 
In addition, noting that in homogeneous, incompressible turbulence
\begin{equation}
\left\langle \delta \uu \cdot \delta \left[ \nabla \left( \frac{u^2}{2} \right) \right] \right\rangle = \left\langle 2 \uu \cdot \nabla \left(\frac{u^2}{2} \right) - \uu' \cdot \nabla \left(\frac{u^2}{2} \right) - \uu \cdot \nabla' \left(\frac{u'^2}{2} \right) \right\rangle =  \left\langle  \nabla \cdot  \left({u^2} \uu \right) + \frac{u^2}{2} \theta' +  \frac{u'^2}{2} \theta  \right\rangle = 0 ,
\end{equation}
one can immediately recover the exact relation for incompressible MHD (equation (27) in \citep{Banerjee17JoP}). Similarly, (\ref{mainhydro}) can be reduced to the corresponding incompressible version, which is identical to equation (9) of \citep{Banerjee17JoP}. Interestingly, in incompressible limit, the total gravitational contribution vanishes identically. \\

\paragraph{\textbf{Effect of mean magnetic field}} 

In incompressible MHD turbulence, it was shown that the effect of a uniform background magnetic field $ \bm B_0$ cancels out  \cite{Banerjee17JoP}. Here we show that in case of compressible turbulence, the effect of mean magnetic field is non-trivial. The magnetic contribution in the exact relation is given by $\pazocal{M}$ where the contribution from $\bb_0 ( \equiv \bm B_0/\sqrt{\mu_0})$ can be separately written as
\begin{eqnarray}
\pazocal {M}_0 = \left\langle \delta \jm \cdot \left( {\bb_0} \times \delta \left(\frac{\jb}{\rho} \right) \right) +  \delta \jb \cdot \left( {\bb_0} \times \delta \uu \right) \right\rangle =  {\bb_0} \left\langle  \delta \rho \left( \uu' \times \frac{\jb}{\rho}  -  \uu \times \frac{\jb'}{\rho'}  \right)  \right\rangle \,  .  \label{background}
\end{eqnarray}
From the above expression we can readily find that, if $\bb_0\ne0$, $\pazocal {M}_0 $ vanishes
\begin{enumerate}[(i)]
\item for incompressible case ($\delta \rho = 0$), or 
\item when $\jm' \times \jb = \jm \times \jb'$, or 
\item if $\left\langle  \delta \rho \left( \uu' \times \frac{\jb}{\rho}  -  \uu \times \frac{\jb'}{\rho'}  \right)  \right\rangle = 0$. \\
\end{enumerate}

\paragraph{\textbf{Effect of alignments}} 

One could consider a hypothetical state where all different sorts of alignments are valid: (a) helical Beltrami flow with $\uu||\bomega$ and $\uu\times\bomega=0$; (b) with a force-free magnetic field, $\jb||\bb$ or $\jb\times\bb=0$, decoupled from the dynamics; (c) with velocity and magnetic field aligned $\uu||\bb$ and $\uu\times\bb=0$ due to the so-called Alfv\'en effect; and (d) with Zeldovich alignment $\uu||\gg$, resulting in cancellation of gravitational terms (if $\uu=\lambda\gg$ with $\lambda=const$). (Note that, a simultaneous presence of all these alignments is hardly realizable.)  One can show that the energy flux rate does not completely vanish and can be written as 
\begin{equation}
\pazocal{Q}= \left\langle \delta \jm \cdot \delta \left[ \nabla \left(\frac{u^2}{2} \right) \right] + \delta \uu \cdot \delta \left[ \left( \nabla \cdot \jm \right) \uu + \rho \, \nabla \left(\frac{u^2}{2} \right)  \right] + \delta \rho \,  \delta \left[ \left( \uu \cdot \nabla \right) e \right]\right\rangle,
\end{equation}
which however vanishes identically for the incompressible case. The residual part contains only kinetic and thermodynamic terms whereas the magnetic and gravitational contributions vanish. This shows that the kinetic relaxation state in compressible turbulence may be quite different from $\uu \parallel \bomega$, even if the compressibility is weak. In fact, the corresponding relaxation state of compressible turbulence would involve the thermodynamic potential energy. Interestingly, one can readily recognise that $\pazocal{Q}$ is nothing but the hydrodynamic contribution for an irrotational flow ($\nabla \times \uu = \bomega= 0$). \\

\paragraph{\textbf{Influence of global rotation}} 

With the current formulation, the effect of a global rotation of the flow field can be handled easily. If we assume that the system is rotating with an angular speed $\bOmega$ (not necessarily constant), then the fluid will experience an additional Coriolis acceleration ${\bm a}_c = ( \uu \times 2 \bOmega)$. The extra flux due to ${\bm a}_c $ is given by 
\begin{equation}
{\pazocal K}_{c} = - 2 \left\langle \delta \jm \cdot \delta \left( \uu \times \bOmega \right) + \delta \uu \cdot \delta \left( \jm \times \bOmega \right) \right\rangle.
\end{equation}
Now, in the case of solid-body rotation, i.e., when $\bOmega$ is a constant, one can easily verify that ${\pazocal K}_{c} $ vanishes, which is similar to incompressible turbulence where also a uniform rotation cannot alter the energy flux rate in the so-called inertial zone. \\

\paragraph{\textbf{Extension to the Hall MHD turbulence}} 

This new exact relation can easily be extended to 3D compressible Hall MHD turbulence. The Hall MHD equations differ from the ordinary MHD equations by the inclusion of the Hall term in the induction equation, which then takes the following form:
\begin{equation}
\p_t \bb = \nabla \times \left( \uu \times \bb \right) - \sqrt{\mu_0} d_i \nabla \times \left(\jb \times \bb \right) + {\bm d_M} ,
\end{equation}
where $d_i$ denotes the ion inertial length of the MHD fluid and $\sqrt{\mu_0} d_i \nabla \times \left(\jb \times \bb \right) $ is the so-called Hall term. In this case, the additional contribution to $\p_t {\pazocal R}$ due to the Hall term can be written as
\begin{equation}
{\pazocal M}_{Hall} = - \frac{\sqrt{\mu_0} d_i}{2} \left\langle \bb' \cdot \nabla \times \left( \jb \times \bb \right) + \bb \cdot  \nabla' \times \left( \jb' \times \bb' \right) \right\rangle =  \frac{\sqrt{\mu_0} d_i}{2} \left\langle \delta \jb \cdot \delta \left( \jb \times \bb \right) \right\rangle.
\end{equation}
The final exact relation for three dimensional, compressible, Hall MHD turbulence is then simply 
\begin{align}
- 4 \varepsilon &= \left\langle  \delta \jm   \cdot   \delta \left[ \left( \uu \cdot \nabla \right) \uu +   \frac{\bb \times \jb}{\rho}  + \gg \right] + \delta \uu \cdot  \delta \left[ \nabla \cdot \left( \jm \otimes \uu \right) 
+  \bb \times \jb  - \rho \gg  \right]  \right\rangle  \nonumber \\
&+  \left\langle  2 \delta \jb\cdot \delta \left[\bb \times \left( \uu - \sqrt{\mu_0} d_i \jb \right) \right]   + \delta \rho \,  \delta \left[ \left( \uu \cdot \nabla \right) e \right] \right \rangle .\label{mainexact1}
\end{align} 
which is considerably simpler than that recently obtained for 3D compressible Hall MHD turbulence in Ref.~\citep{Andres17Hall}. \ADD{Note that, in MHD models of star forming clouds based on single-fluid approximation, the ambipolar diffusion is found to be the dominant mechanism at small scales \citep{Tassis04} and the turbulent energy decay is found to be hardly affected by the Hall terms \citep{Matthaeus03}. However, a number of studies \citep{Wardle99, Wardle04} have shown that Hall diffusion can be significant in dense regions of molecular clouds with particle density ranging from $(10^7- 10^{11})$~cm$^{-3}$. Moreover, using multi-fluid models, it was also found \citep{Downes08} that the Hall effect can be important for the small-scale turbulence in molecular clouds under a broad range of conditions. Here, the inclusion of Hall terms makes the relation more complete, which allows one to formulate a number of reduced versions (e.g., including gravity without Hall effect or including compressible Hall MHD turbulence without gravity, etc.).}

\section{Discussion} 

In this current work, possibly for the first time, an exact relation has been derived for three-dimensional, compressible turbulence for a self-gravitating, magnetized isothermal fluid, which will be of interest to both space physicists and astrophysicists. In contrast with the previously obtained exact relations for compressible turbulence \citep{Galtier11, Banerjee13, Banerjee14, Andres17, Banerjee17, Andres17Hall}, here we use an alternative formulation and show that the final result expressed in terms of two-point differences is notably more straightforward. \ADD{This relation provides an opportunity to obtain accurate estimates of the turbulent heating rate of the solar wind for a general anisotropic case. However, the evaluation of terms such as $(\uu \cdot \nabla) \uu$ will require the use of multi-spacecraft data (e.g., Cluster, Themis, MMS, etc.).}  From numerical point of view, it is also easier to calculate all the terms of the right-hand side. In fact, the general form of the relation is almost identical to that of the incompressible ones, thereby justifying a more universal applicability for turbulence, compared to the previous flux-source formulation. By a simple calculation, we showed that, similar to incompressible turbulence, a solid-body rotation cannot alter the energy flux rate in compressible turbulence. Moreover, the inclusion of the Hall term does not alter the global form of the equation either. We also showed that the presence of a uniform background magnetic field has a non-trivial effect on the energy flux in compressible turbulence. Finally it was also shown that, unlike in incompressible turbulence, the presence of alignments does not make the contributions from the kinetic and the thermodynamic potential energies vanish. At the same time, the magnetic (without the Hall term) and the gravitational contributions would vanish under aligned condition.

\begin{acknowledgments}
The work of A.K. was supported, in part, by the National Science Foundation Grant No.~AST-141227.
\end{acknowledgments}



\end{document}